\documentclass[prc, twocolumn, superscriptaddress, preprintnumbers, amsmath, amssymb]{revtex4}

\allowdisplaybreaks
\allowdisplaybreaks[4]

\usepackage{graphicx}% Include figure files
\usepackage{dcolumn}% Align table columns on decimal point
\usepackage{bm}% bold math
\usepackage{multirow}

\usepackage{color}%,mathptmx,helvet}
\usepackage[
dvipdfm,
            pdfstartview=FitH,
            CJKbookmarks=true,
            bookmarksnumbered=true,
            bookmarksopen=true,
            colorlinks=true,
            pdfborder=001,
            linkcolor=blue,
            anchorcolor=blue,
            citecolor=blue
            ]{hyperref}

\begin{document}

\title{Transverse wobbling in an even-even nucleus}

\author{Q. B. Chen}\email{qbchen@pku.edu.cn}
\affiliation{Physik-Department, Technische Universit\"{a}t
M\"{u}nchen, D-85747 Garching, Germany}

\author{S. Frauendorf}\email{sfrauend@nd.edu}
\affiliation{Physics Department, University of Notre Dame, Notre
Dame, IN 46556, USA}

\author{C. M. Petrache}\email{petrache@csnsm.in2p3.fr}
\affiliation{Centre de Sciences Nucl\'eaires et Sciences de la
Mati\`ere, CNRS/IN2P3, Universit\'{e} Paris-Saclay, B\^at.~104-108,
91405 Orsay, France}

\date{\today}

\begin{abstract}

Two new bands built on the two-quasiparticle $\pi(h_{11/2})^2$
configuration of the even-even nucleus $^{130}$Ba are investigated
using constrained triaxial covariant density functional theory
combined with quantum particle rotor model calculations. The energy
difference between the two bands, as well as the available
electromagnetic transition probabilities $B(M1)_{\textrm{out}}/B(E2)_{\textrm{in}}$
and $B(E2)_{\textrm{out}}/B(E2)_{\rm{in}}$, are well reproduced. The
analysis of the angular momentum geometry reveals that the higher band
represents transverse wobbling motion of a two-quasiparticle configuration.
This is the first example of two-quasiparticle wobbling bands in an
even-even nucleus.

\end{abstract}

\maketitle

%%%%%%%%%%%%%%%%%%%%%%%%%%%%%%%%%%%%%%%%%%%%%%%%%%%%%%%%%%
%                    begin  introduction
%%%%%%%%%%%%%%%%%%%%%%%%%%%%%%%%%%%%%%%%%%%%%%%%%%%%%%%%%%

The wobbling motion, proposed by Bohr and Mottelson~\cite{Bohr1975},
is an unique feature of triaxially deformed rotating nuclei. Its
analog in classical mechanics is the motion of a free asymmetric
top. Uniform rotation about the principal axis with the largest
moment of inertia has the lowest energy for given angular momentum.
At slightly larger energy this axis executes harmonic precession
oscillations about the space-fixed angular momentum vector. The
corresponding quantal energy spectrum is a series of rotational
$\Delta I=2$ bands where the signature of the bands alternates with
the increasing number of oscillation quanta $n$. The $\Delta I=1$,
$n\rightarrow n-1$ $E2$ transitions are collectively enhanced.

The predicted wobbling mode for even-even nuclei has not been
experimentally confirmed yet. Instead wobbling bands have been
reported in the odd-$A$ mass nuclei where intrinsic angular momentum
is involved. Most of them are in odd-proton nuclei:
$^{161}$Lu~\cite{Bringel2005EPJA}, $^{163}$Lu~\cite{Odegaard2001PRL,
Jensen2002PRL}, $^{165}$Lu~\cite{Schonwasser2003PLB},
$^{167}$Lu~\cite{Amro2003PLB}, and $^{167}$Ta~\cite{Hartley2009PRC}
in $A\approx 160$ mass region and $^{135}$Pr~\cite{Matta2015PRL,
Sensharma2019PLB} in the $A\approx 130$ mass region. Very recently,
a wobbling band was also reported in the odd-neutron nucleus
$^{105}$Pd~\cite{Timar2019PRL} in the $A\approx 100$ mass region.

So far wobbling mode has only been observed in odd-$A$ nuclei. The
energy difference between the wobbling bands has been found to
decrease with increasing spin, contrary to the behavior expected for
even-even nuclei~\cite{Bohr1975}. Frauendorf and D\"onau~\cite{Frauendorf2014PRC}
interpreted this behavior as the consequence of the perpendicular orientation
of the odd-particle angular momentum to the axis with the maximal moment of
inertia. They called this coupling \textit{transverse} wobbling (TW) to
distinguish it from the alternative coupling scheme which corresponds
to parallel orientation of the odd particle angular momentum to
the axis with the maximal moment of inertia. For this \textit{longitudinal}
wobbling mode (LW) the wobbling energy increases with spin.

The evidence for wobbling in even-even nuclei is fragmentary. For
instance, Refs.~$^{112}$Ru~\cite{Hamilton2010NPA, Frauendorf2014PRC} and
$^{114}$Pd~\cite{Y.X.Luo2013proceeding} interpreted the ``$\gamma$-bands"
as $n=1$ and $2$ wobbling bands, because the energy of their odd-spin
members lies below the mean energy of the adjacent even-spin levels.
Such a fingerprint is considered as evidence of triaxial
deformation~\cite{Zamfir1991PLB} (see a detailed discussion
in Ref.~\cite{Frauendorf2015IJMPE}), for which the wobbling mode
develops with increasing spin (c.f.~\cite{Bohr1975, Frauendorf2014PRC}).
However no electromagnetic transition data were reported to put the
wobbling interpretation on solid ground. At present, the appearance of
wobbling modes is not yet demonstrated for even-even nuclei.

Very recently, in Ref.~\cite{Petrache2019PLB}, a large variety of
band structures has been reported in the even-even nucleus
$^{130}$Ba, out of which, a pair of bands with even and odd spins,
labeled S1 and S1$^\prime$, attracted our attention. According to
the quasiparticle alignment analysis, the bands are built on two
rotational aligned proton $h_{11/2}$ particles. Such a configuration
can fulfill the conditions for TW motion.

In this paper, we report a theoretical investigation which suggests
that the two newly observed two-quasiparticle bands S1 and S1$^\prime$
of $^{130}$Ba can be interpreted as the zero- and one-phonon states
of the TW mode. This is the first example of wobbling motion
based on a two-quasiparticle configuration. The analysis of the data shows that
the TW regime is much more stable than for the known cases with one odd quasiparticle.
That is, we present the best known case for TW.

The experimental information relevant for the present work has been
recently reported in Ref.~\cite{Petrache2019PLB}, in which the
mixing ratios $\delta$ of the transitions connecting the two
wobbling candidates have been extracted from the analysis of the
angular distribution of the $\gamma$-rays emitted by the excited
$^{130}$Ba nucleus. The mixing ratios were subsequently used to
deduce the $B(M1)_{\textrm{out}}/B(E2)_{\textrm{in}}$ and
$B(E2)_{\textrm{out}}/B(E2)_{\textrm{in}}$ ratios of reduced
transition probabilities for the $\Delta I=1$ transitions connecting
band S1$^\prime$ to band S1. The experimental data were obtained
from a fusion-evaporation reaction in  which $^{130}$Ba was
populated via the $^{122}$Sn($^{13}$C,5n) reaction at a beam energy
of 65 MeV. The mixing ratios $\delta$ of the $M1/E2$ transitions
were deduced from the transition intensities measured at the four
angles available in the GALILEO array~\cite{Dobon2014AR,
Testov2019proceeding} by employing a method developed by
Matta~\cite{Matta2015PRL} for the analysis of angular-distribution
measurements. For many transitions there are two solutions for
$\delta$ in the $\chi^2$ plot, with the absolute values larger than
1, and smaller than 1. For all transitions analyzed in the present
work, the $\delta$ values smaller than 1 have been adopted, since
they have smaller $\chi^2$ values. However, one cannot completely
exclude the larger values based only on the angular distribution
measurement. On the other hand, such larger values would reinforce
the conclusions of the present work. The resulting values are
collected in Table~\ref{tab1}. More experimental details can be
found in Refs.~\cite{Y.H.Qiang2019PRC, Petrache2019PLB}.

To determine the shape, we carried out triaxial constrained
covariant density functional theory (CDFT) calculations with the
effective interaction PC-PK1~\cite{P.W.Zhao2010PRC}. The details are
given in Refs.~\cite{J.Meng2006PRC, J.Meng2016book}. The
configuration $\pi(h_{11/2})^2$ was assigned to bands S1 and
S1$^\prime$, which was fixed by  requiring that the overlap between
the Slater determinants of adjacent points on the deformation grid
had to be larger than 0.9. The minimum was found at the triaxial
shape with $(\beta=0.24, \gamma=21.5^\circ)$ and an excitation
energy of 3.13 MeV with respect to the ground state $(\beta=0.23,
\gamma=14^\circ)$, which favorably compares with the experimental
excitation energy of 3.79 MeV of the $I$=10 state of band S1. To
check the angular momentum dependence of the shape, we carried out
(tilted axis cranking) TAC calculations~\cite{Frauendorf2000NPA} assuming
zero proton pairing and varying neutron paring. The deformation changes are
minor and not very sensitive to the neutron pairing. For $I=14
\rightarrow 24$ we found that $(\beta,\gamma)=(0.23,30^\circ
)\rightarrow(0.20,26^\circ )$ for $\Delta_n=0$ and $(0.21,24^\circ
)\rightarrow(0.20,29^\circ )$ for  $\Delta_n=0.8$ MeV. At fixed
deformation, the dynamic moment of inertia ${\cal J}^{(2)}=d I/d
\omega$ is about constant for $\Delta_n=0$ and increases for
$\Delta_n>0$, where the rate sensitively depends on $\Delta_n$.

We performed particle rotor model (PRM) \cite{Hamamoto2002PRC,
Frauendorf2014PRC, W.X.Shi2015CPC, Streck2018PRC} calculations for
the adopted $\pi (h_{11/2})^2$ configuration and the CDFT
deformation parameters $(\beta=0.24, \gamma=21.5^\circ)$. The
protons are described by a single-$j$ shell
Hamiltonian~\cite{Ring1980book} assuming zero pairing. The triaxial
rotor is parametrized by the three spin-dependent moments of inertia
$\mathcal{J}_i=\Theta_i(1+cI)$, in which $i=s$, $m$, $l$ denotes the
short, medium, and long axes, respectively. The parameters
$\Theta_s$, $\Theta_m$, $\Theta_l$=1.09, 1.50,
0.65~$\hbar^2/\textrm{MeV}$ and $c=0.59$ are determined by adjusting
the PRM energies to the experimental energies of the zero- and
one-phonon bands. The spin-dependence is attributed to the presence
of neutron pairing. The fitted ratios
$\mathcal{J}_s$/$\mathcal{J}_m$/$\mathcal{J}_l$=0.73/1.00/0.43,
different from the ratios 0.40/1.00/0.14 for the irrotational flow
type moments of inertia~\cite{Ring1980book}. The larger ratio
$\mathcal{J}_s$/$\mathcal{J}_m$ stabilizes the TW mode. Microscopic
cranking calculations give the larger ratio
$\mathcal{J}_s$/$\mathcal{J}_m=0.60/1.00$~\cite{Frauendorf2018PRC}.
One also has to take into account that the deformation represents an
average value, around which substantial fluctuations exist. These
reduce $\mathcal{J}_m$ from its irrotational flow value for the
average $\gamma$ value (see Fig. 1 of~\cite{Frauendorf2018PRC}).
Modifying the deformation parameters within the range found by the
TAC calculations and keeping $\mathcal{J}_i$ the same did not change
the results in a substantial way. For the electromagnetic
transitions we used the intrinsic quadrupole moments
$Q_0=Q\cos\gamma$ and $Q_2=Q\sin\gamma/\sqrt{2}$ with
$Q=(3/\sqrt{5\pi})R_0^2Z\beta$, $R_0=1.2A^{1/3}$ fm and the
gyromagnetic ratios  $g_R=Z/A=0.43$ for the rotor and
$g_\pi (h_{11/2})= 1.21$ for the protons.

\begin{figure}[!ht]
  \begin{center}
    \includegraphics[width=6.5 cm]{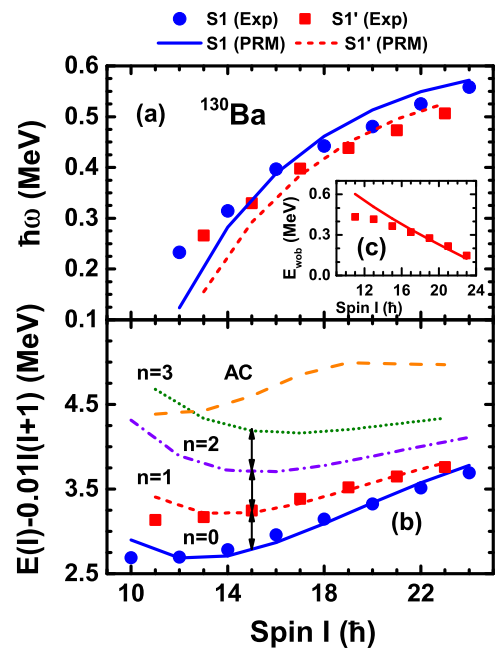}
    \caption{Energetics of the lowest bands based on the $\pi(h_{11/2})^2$ in $^{130}$Ba.
    Panel (a) Experimental and PRM rotational frequency for the bands S1 and
    S1$^\prime$. Panel (b) PRM energies minus a common rigid rotor reference.
    The bands based on the lowest $\pi(h_{11/2})^2$ configuration AB are labelled by the
    wobbling phonon number $n$ assigned to them. The $n=0$ band based on the excited
    $\pi(h_{11/2})^2$ configuration is labelled by AC. The known experimental energies are
    included. Inset (c): Wobbling energies $E_{\textrm{wob}}(I)=E_{n=1}(I)-[E_{n=0}
    (I+1)+E_{n=0}(I-1)]/2$ obtained by PRM compared with the experimental values
    obtained from bands S1 ($n=0$) and S1$^\prime$ ($n=1$).
   }\label{fig1}
  \end{center}
\end{figure}

The calculated rotational frequency $\hbar\omega(I)=[E(I)-E(I-2)]/2$ and
energy $E(I)$ spectra for bands S1 and S1$^\prime$, in comparison
with the experimental data, are shown in Fig.~\ref{fig1}. It is seen that
the PRM calculations reproduce well the bands S1 and S1$^\prime$.
In agreement with the experimental observation, the calculated wobbling
energy $E_{\textrm{wob}}(I)$ decreases in the whole spin region, providing
evidence for TW motion.

The experimental and theoretical mixing ratios $\delta$, as well as
the transition probability ratios $B(M1)_{\textrm{out}}/B(E2)_{\textrm{in}}$
and $B(E2)_{\textrm{out}}/B(E2)_{\textrm{in}}$ for the transitions from
band S1$^\prime$ to S1 are compared in Table~\ref{tab1}. The
calculated $B(E2)_{\textrm{out}}/B(E2)_{\textrm{in}}$ values agree
with experiment within the uncertainties. The
$B(E2)_{\textrm{out}}/B(E2)_{\textrm{in}}$ is proportional to $\tan^2\gamma$~\cite{Bohr1975}.
The PRM calculations using $\gamma=21.5^\circ$ can reproduce the experimental
$B(E2)_{\textrm{out}}/B(E2)_{\textrm{in}}$ value. Thus, the
microscopic input of the triaxial deformation parameter from the
CDFT calculation is appropriate. The large ratios indicate that the
$E2$ transitions from S1$^\prime$ to S1 are highly collective. This
is the fingerprint of TW, which represents a wobbling of the
triaxial charge density with respect to the angular momentum vector.
The theoretical $B(M1)_{\textrm{out}}/B(E2)_{\textrm{in}}$ ratios
are somewhat on the large side compared to experiment.

\begin{table}[!ht]
\caption{The experimental and theoretical mixing ratios $\delta$ as
well as the transition probability ratios
$B(M1)_{\textrm{out}}/B(E2)_{\textrm{in}}$ and
$B(E2)_{\textrm{out}}/B(E2)_{\textrm{in}}$ for the transitions from
band S1$^\prime$ to band S1 of $^{130}$Ba.} \label{tab1}
\begin{tabular}{cccccccccccc}
\toprule
\multirow{2}*{$I$ ($\hbar$)}
 & \multicolumn{2}{c}{$\delta$} &
& \multicolumn{2}{c}{$\frac{B(M1)_{\textrm{out}}}{B(E2)_{\textrm{in}}}(\frac{\mu_N^2}{e^2b^2})$} &
& \multicolumn{2}{c}{$\frac{B(E2)_{\textrm{out}}}{B(E2)_{\textrm{in}}}$} \\
\cline{2-3}
\cline{5-6}
\cline{8-9}
 & Exp  & ~PRM~ & & Exp  & ~PRM~ & & Exp  & ~PRM~ \\
\hline
$13$ &  $-0.58^{+13}_{-13}$  & $-0.67$ && $0.36^{+19}_{-13}$ & 1.11 && $0.32^{+18}_{-15}$ & 0.51\\
$15$ &  $-0.62^{+10}_{-10}$  & $-0.68$ && $0.38^{+61}_{-16}$ & 0.90 && $0.36^{+70}_{-19}$ & 0.42\\
$17$ &  $-0.62^{+10}_{-10}$  & $-0.68$ && $0.23^{+22}_{-09}$ & 0.76 && $0.22^{+27}_{-10}$ & 0.35\\
$19$ &  $-0.60$              & $-0.66$ && $0.25^{+23}_{-08}$ & 0.67 && $0.22^{+21}_{-07}$ & 0.29\\
$21$ &  $-0.60$              & $-0.63$ && $0.43^{+35}_{-13}$ & 0.63 && $0.41^{+34}_{-13}$ & 0.25\\
\hline
\end{tabular}
\end{table}

Note that in comparison with TW in odd-$A$ nuclei, as $^{135}$Pr~\cite{Matta2015PRL,
Sensharma2019PLB} and $^{105}$Pd~\cite{Timar2019PRL}, the
$B(M1)_{\textrm{out}}/B(E2)_{\textrm{in}}$ in $^{130}$Ba is about three
times larger. This is attributed to the fact that one more high-$j$
quasiparticle is involved in the configuration of the S1, S1$^\prime$ bands,
which enlarges the $M1$ matrix elements. As a consequence,
the relative contribution of the $E2$ component to the  $\Delta I=1$
transitions [calculated as $\delta^2/(1+\delta^2)$] is smaller
($\sim 25\%$) and does not dominate like in the case of the
one-quasiparticle bands of the odd-even wobbling nuclei.

The overestimation of the $B(M1)_{\textrm{out}}$ has already been
observed in the PRM calculations for $^{135}$Pr~\cite{Frauendorf2014PRC,
Matta2015PRL} and the Lu isotopes~\cite{Frauendorf2014PRC}. According
to the quasiparticle-random-phase approximation (QRPA)
calculations~\cite{Frauendorf2015PRC}, the wobbling motion is
not a pure orientation vibration of the quadrupole mass tensor
with respect to the angular momentum vector, but also coupled
to the vibrations of the proton and neutron currents against
each other, i.e., the scissor mode. The coupling to the scissor
draws $M1$ strength from the TW mode, which is not taken into in
the PRM.

\begin{figure}[!ht]
  \begin{center}
    \includegraphics[width=\linewidth]{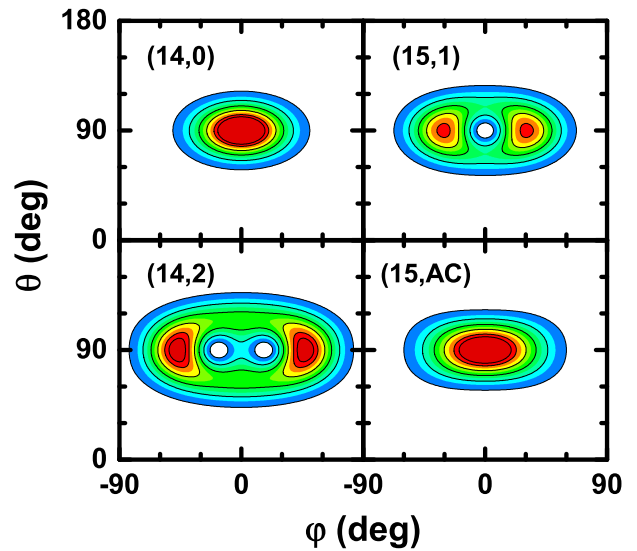}
    \caption{Distributions of the probability  $\mathcal{P}(\theta,\varphi)$ for the orientation
    of the angular momentum $\bm{J}$ with respect to the body-fixed frame (``azimuthal plots").
    Brown indicates maximal and white minimal probability.
    Contour lines show equal differences of the probability  density.
    Here, $\theta$ is the angle between $\bm{J}$
    and the $l$-axis, and $\varphi$ is the angle between the projection of $\bm{J}$ onto the
    $sm$-plane and the $s$-axis. Due to the $\textrm{D}_2$ symmetry, $\mathcal{P}(\theta,\varphi)$
    is an even function of $\varphi$ with a period of $\pi$. In accordance with Fig. \ref{fig1},
    the panels are labelled by  $(I,n$).}\label{fig2}
  \end{center}
\end{figure}

\begin{figure}[!ht]
  \begin{center}
    \includegraphics[width=8.5 cm]{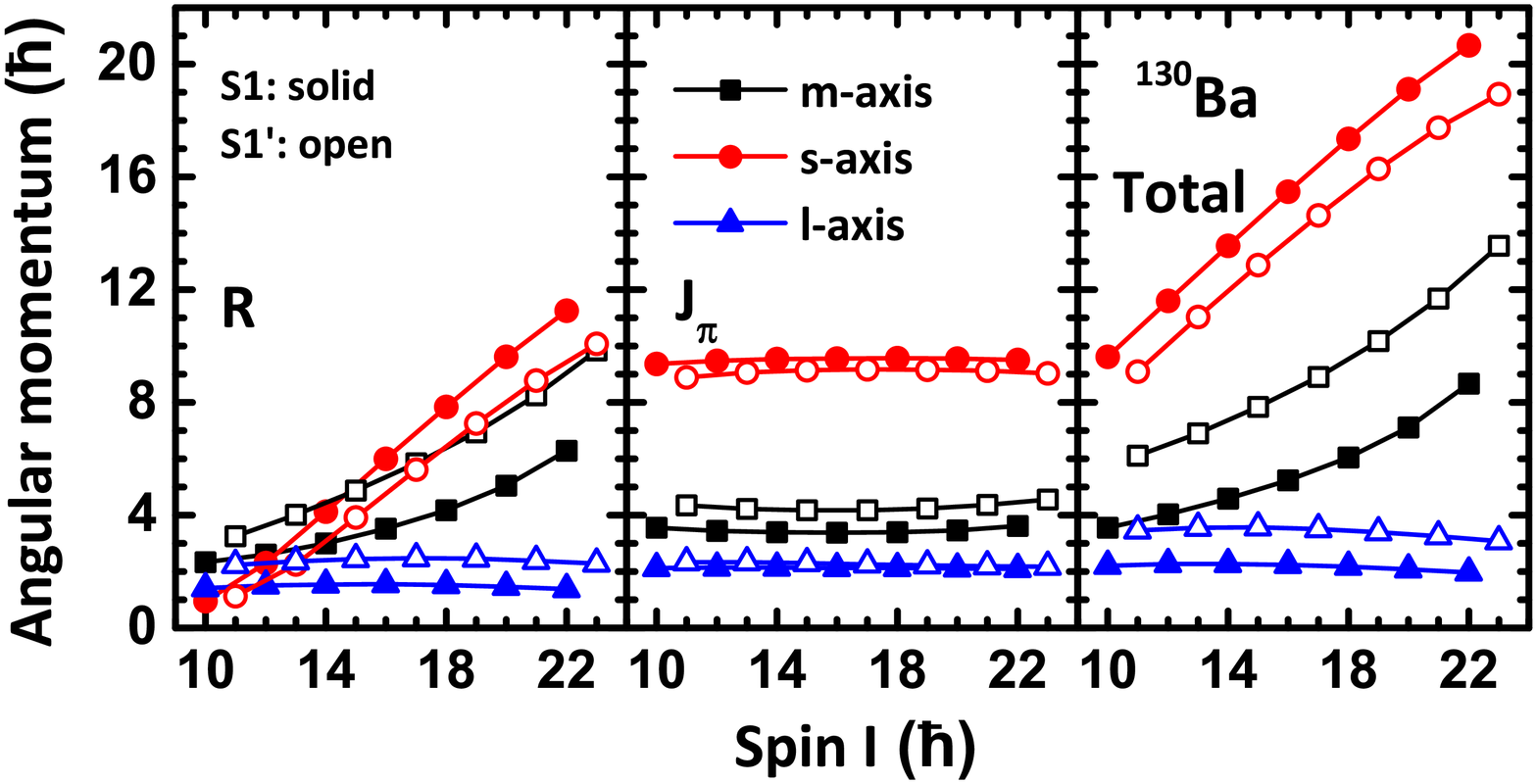}
    \caption{The root mean square angular momentum components along the medium
    ($m$-, squares), short ($s$-, circles), and long ($l$-, triangles) axes of
    the rotor $\bm{R}$, valence protons $\bm{J_\pi}$, and total angular momentum $\bm{I}$   for the bands S1
    (full symbols) and S1$^\prime$ (open symbols).}
    \label{fig3}
  \end{center}
\end{figure}

Fig.~\ref{fig2} shows the probability density distributions
$\mathcal{P}(\theta,\varphi)$ for the orientation of the angular
momenta $\bm{J}$ with respect to the body-fixed
frame~\cite{F.Q.Chen2017PRC, Q.B.Chen2018PRC_v1, Streck2018PRC,
Q.B.Chen2019tbp} at spin $I=14$ and 15, respectively.
Fig.~\ref{fig3} shows  the root mean square components along the
medium ($m$-), short ($s$-), and long ($l$-) axes of the $\bm{R}$,
of the valence protons $\bm{J_\pi}$, and of the total angular
momentum $\bm{I}$, for the bands S1 and S1$^\prime$.

The figures corroborate the scenario of stable TW. The distributions
$\mathcal{P}(\theta,\varphi)$ are centered with the
$\theta=90^\circ$ plane, corresponding to the very small
$l$-component of $\bm{I}$ in Fig.~\ref{fig3}. The $n=0$ state has a
maximum at $\varphi=0^\circ$ corresponding to the maximal alignment
of $\bm{I}$ with the $s$- axis allowed by quantum mechanics. The
$n=1$ state  has a minimum there. The maximal probabilities lie on a
rim revolving the minimum, which reflects the wobbling motion
(precession) of $\bm{I}$ about the $s$-axis. These are the expected
distributions for the $\varphi$-symmetric $n=0$  and
$\varphi$-antisymmetric $n=1$ states.  The fact that the
distributions centered at $\varphi=0^\circ$ and $\varphi=\pm
180^\circ$ do not merge at $\varphi=\pm 90^\circ$ indicates that the
TW mode is stable.

Fig. \ref{fig3} illustrates how the rms components of $\bm{R}$,
$\bm{J_\pi}$, and $\bm{I}$ change with $I$. The $\bm{J_\pi}$ of
valence proton particles tends to align with the $s$-axis, which
corresponds to maximal overlap with the triaxial
core~\cite{Frauendorf1996Z.Phys.A}. The $s$-component is constant
9.5 and 9$\hbar$ for S1 and S1$^\prime$, respectively. The
transverse geometry is more stable than for the one-proton bands,
where a tilt of the proton angular momentum toward the $m$-axis
appears~\cite{Streck2018PRC, Q.B.Chen2019tbp}.

The $l$-component of $\bm{R}$ stays small because the moment of
inertia for rotation around the $l$-axis is the smallest. In order
to lower the energy $\bm{R}$ favors the $sm$-plane. For band S1, it
increases more along the $s$-axis than along the $m$-axis. Combined
with  $\bm{J_\pi}$, the $s$-component of $\bm{I}$ is larger than the
$m$-component. For band S1$^\prime$, the $s$- and $m$-components of
the rotor are about the same. This is attributed to the fact that
the transverse wobbling excitation is achieved by adding rotor
angular momentum component along the $m$-axis~\cite{Streck2018PRC}.
The instability point of TW, above which $\bm{I}$ increases  by
adding $\bm{R}$ along the $m$-axis, is not reached, which is
consistent with the continued down trend of the wobbling frequency
seen in Fig.~\ref{fig1}.

As suggested by the labeling in Fig.~\ref{fig1}, the next higher bands
are interpreted  them as the $n=2$ and $n=3$  wobbling excitations.
In particular for $12\leq I\leq18$, the approximately equal energy
spacing provokes the multi-phonon interpretation. In accordance with
this interpretation, Fig.~\ref{fig2} shows for the $n=2$ band
a rim of  $\mathcal{P}(\theta=90^\circ,\varphi)$ revolving $\varphi=0$.
Its distance to the origin is lager, which reflects the wider precession cone of the $n=2$
wobbling excitation. The maximum at $\varphi=0$ between two symmetrically
located minima is the topology of a $\varphi$-symmetric wave
function with two nodes. The distribution $\mathcal{P}(\theta,\varphi)$
for the $n=3$  band (not shown) has yet wider rim and a minimum at
$\varphi=0^\circ$ between two symmetrically located maxima and two
symmetrically located minima further out, which is the
topology of a $n=3$ wave function with three nodes.

\begin{figure}[!ht]
  \begin{center}
    \includegraphics[width=5.5 cm]{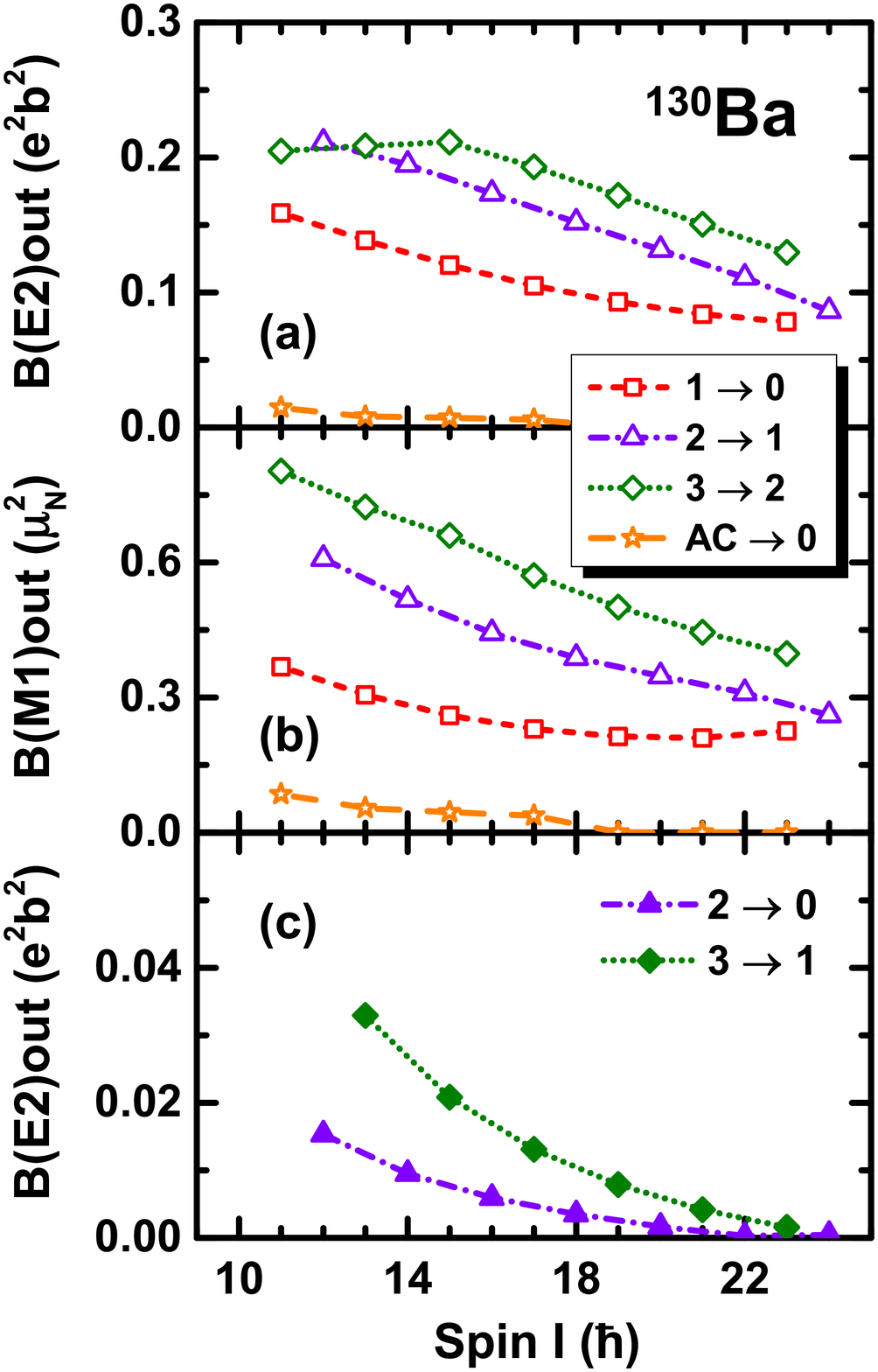}
    \caption{The calculated $B(E2)$ and $B(M1)$ values for inter band transitions:
    (a) $B(E2)_{\textrm{out}}(n, I \to n-1, I-1)$, (b) $B(M1)(n, I \to n-1, I-1)$,
    (c) $B(E2)_{\textrm{out}}(n, I \to n-2, I-2)$.}\label{fig4}
  \end{center}
\end{figure}

The calculated $B(E2)_{\textrm{out}}$ and $B(M1)_{\textrm{out}}$ values
for the $n=1$, 2, and 3 bands are shown in Fig.~\ref{fig4}. As expected,
the $B(E2, n\rightarrow n-1)_{\textrm{out}}$ and $B(M1, n \rightarrow n-1)_{\textrm{out}}$
values increase with the phonon number $n$, however more slowly than
the harmonic limit $\propto n$. For a simple triaxial rotor the harmonic
limit is approached with increasing angular momentum, because the
amplitude of the wobbling motion $\propto 1/\sqrt {I}$~\cite{Bohr1975}.
For the TW mode this trend is counteracted by the approach of the
critical angular momentum for instability. The transitions
$n \rightarrow n-2$, which are forbidden in the harmonic limit,
are strongly reduced compared with the allowed ones.

Adopting cranked shell model terminology, we call the lowest three
$h_{11/2}$ proton orbitals in the rotating potential A, B, C,  which
have the respective signatures $\alpha=-1/2$, $1/2$, $-1/2$. Then S1
is the $\alpha=0$ configuration [AB]. The two protons may combine to
the $\alpha=1$ odd-$I$ configuration [AC]. Fig.~\ref{fig2} shows
that $\bm{I}$ is aligned with the short axis, similar to [AB]. As seen
in Fig.~\ref{fig1}, the AC band lies about 2 MeV above the AB band S1.
Fig.~\ref{fig4} shows that the $B(E2)_{\textrm{out}}$ and $B(M1)_{\textrm{out}}$
values that connect the AC band with S1 are much smaller than the ones that
connect the $n=1$ wobbling excitation S1$^\prime$ with S1. This is
analogous to the case of $^{135}$Pr, where strong
$B(E2)_{\textrm{out}}$ and $B(M1)_{\textrm{out}}$ values connect the
$n=1$ wobbling band with the yrast band whereas the signature
partner band has very weak connecting transitions that are not seen
in experiment~\cite{Matta2015PRL}.

In summary, the pair of newly observed bands S1 and S1$^\prime$
built on the $\pi(h_{11/2})^2$ configuration in $^{130}$Ba were
investigated combining triaxial CDFT with quantum PRM
calculations. The experimental energy spectra, energy difference
between the two bands, as well as the available electromagnetic
transition probabilities $B(M1)_{\textrm{out}}/B(E2)_{\textrm{in}}$
and $B(E2)_{\textrm{out}}/B(E2)_{\rm{in}}$ are well reproduced. The
collective enhancement of the inter-band $E2$ transitions is clear
evidence for the wobbling character of the higher band. The detailed
analysis of the angular momentum geometry demonstrates the stable TW
character of the excited bands, which makes it the first example for
TW based on a two-quasiparticle configuration in an even-even
nucleus and the most stable TW case observed so far.

\begin{acknowledgments}

This work was supported  by Deutsche
Forschungsgemeinschaft (DFG) and National Natural Science Foundation
of China (NSFC) through funds provided to the Sino-German CRC 110
``Symmetries and the Emergence of Structure in QCD'' (DFG Grant
No. TRR110 and NSFC Grant No. 11621131001), and U.S. DoE Grant
No. DE-FG02-95ER4093.

\end{acknowledgments}

%%%%%%%%%%%%%%%%%%%%%%%%%%%%%%%%%%%%%%%%%%%%%%%%%%%%%%%%
%                  begin refereee
%%%%%%%%%%%%%%%%%%%%%%%%%%%%%%%%%%%%%%%%%%%%%%%%%%%%%%%%

\end{document}